\documentclass[preprint,showpacs,preprintnumbers,amsmath,amssymb,prb,groupedaddress]{revtex4}
\usepackage{graphicx}
\usepackage{dcolumn}
\usepackage{bm}
\usepackage{booktabs}
\usepackage{multirow}
\usepackage{color}
\begin{document}
\title{{Thermoelectric efficiency   of nanoscale devices in the linear regime}  }

\author{G. Bevilacqua$^{1}$,  G. Grosso$^{2,3}$, G. Menichetti$^{2,3}$ and G. Pastori Parravicini$^{2,4}$}
\affiliation{$^1$ DIISM, Universit\`{a} di Siena, Via Roma 56, I-53100 Siena, Italy}
\affiliation{$2$ Dipartimento di Fisica ``E. Fermi'', Universit\`{a} di Pisa, Largo Pontecorvo 3, I-56127 Pisa, Italy}
\affiliation{$3$ NEST, Istituto Nanoscienze-CNR, P.za San Silvestro 12, I-56127 Pisa, Italy}
\affiliation{$4$ Dipartimento di Fisica ``A. Volta'', Universit\`{a} di Pavia, Via A. Bassi, I-27100 Pisa, Italy}

\date{\displaydate{date}}
\date{\today}
\begin{abstract}
We study quantum transport through two-terminal nanoscale 
    devices in contact with two  particle reservoirs at different 
    temperatures and chemical potentials.  We  discuss  the general 
    expressions controlling  the electric charge current,  heat currents and 
    the efficiency of energy transmutation in steady conditions  
    in the linear regime. 
    With focus in the parameter domain where the electron system  acts  
    as a  power-generator,  we  elaborate workable   
     expressions for optimal efficiency   and thermoelectric parameters 
     of nanoscale devices.   
    The general  concepts   are set at work  in the paradigmatic cases  
    of  Lorentzian resonances and antiresonances, 
    and the encompassing Fano  transmission function: the  treatments 
    are fully  analytic, in terms  of the   trigamma functions and Bernoulli numbers.  
    From the general curves here reported describing transport through the above model transmission functions,
    useful guidelines  for optimal efficiency and thermopower can be inferred for engineering  nanoscale devices in  energy regions where they show similar transmission functions.

 \end{abstract}
 \pacs{84.60.Bk	,85.80.Fi,65.80.-g,73.23.Ad}

\maketitle

\section{INTRODUCTION}
Thermoelectricity  is an old and young subject of enormous interest both for the fundamental physical phenomena involved~\cite{DRESS12,ZLATIC14}  and the technological  applications.\cite{SALVO99}

At the birth of research  of  thermoelectric (TE) materials, Seebeck demonstrated that it is possible to convert waste heat into electricity,  while Peltier showed that refrigeration of a TE material can be obtained pumping heat by means of electricity. After almost two centuries, it is still a central problem to find the conditions to realize a most efficient Carnot machine for a given finite power output~\cite{WHIT14} also in conditions of large temperature  and electrical potential gradients.\cite{MUTTALIB13}

The energy conversion efficiency of a TE material is measured by the figure of merit dimensionless number  defined as $ZT=\sigma S^2 T/(\kappa_{el}+\kappa_{ph})$, where $\sigma$ is the electronic conductance, $S$ the Seebeck  coefficient, $T$ the absolute temperature and $\kappa_{el}$ ($\kappa_{ph}$) is the electronic (phononic) contribution to the thermal conductance. 

The promise of a TE material with highest figure of merit is a challenge for theoretical and experimental research.
\cite{DRESS12} At first sight the  way to maximize $ZT$ for a given material could seem to increase the quantity $\sigma S^2$, for instance enhancing the charge carriers density by doping, or reducing the contributions to its thermal conductance.
However, increasing $\sigma$ (or $S$) without increasing $\kappa_{el}$, is a conflicting task and still remains the goal: in fact room temperature values of $ZT$ for the best bulk TE materials are around unity, in a range of values not yet satisfactory for large-scale applications.

An alternative approach~\cite{SOFO96}  was suggested  by Mahan and Sofo in 1996.
 Starting from a given phononic thermal conductivity of a TE material, and the expression of the transport coefficients given by the Boltzmann equation, 
they looked for the electronic structure which generates  an energy dependent transport distribution function able to maximize the figure of merit.
 Their mathematical approach led  to conclude that a delta-shaped transport distribution function maximize the transport properties.  Successive contributions~\cite{FAN11,JEONG12} addressing the effect of more realistic  band structure and transmission shapes  evidenced that finite band-widths (e.g. of rectangular shape)  produce higher thermoelectric performances and this occurs both in the linear~\cite{CHANG14} and nonlinear~\cite{WHIT14} regime.

The concept of engineering of the electronic band structure to enhance the figure of merit received  great impulse from progress in nanotechnology~\cite{DRESS07}
 and advances in the synthesis of complex~\cite{SNYDER08} and organic materials.\cite{KATZ12}
Modulation of the electronic properties of nano- and of organic molecular-electronic  materials have opened perspectives for the control  and enhancement of $ZT$, mainly due to confinement effects and the possibility they offer to reduce the phononic thermal conductivity.\cite{YU10,LI14} In particular, the prediction~\cite{BERG09, BERG10} of giant thermoelectric effects on conjugated single molecule junctions characterized by nodes and supernodes in the transmission spectrum contributed to increase the interest  toward organic thermoelectrics. 

In the present paper we focus on a general two-terminal nanoscale device in contact with two particle reservoirs, the left and right ones, at different temperatures and chemical potentials:   $T_L, \mu_L$  and    $T_R, \mu_R$. 
 The general expressions provided by the Keldysh formalism~\cite{DATTA,GOOD09,CUEVAS10,BALZER13,CRESTI06} are the most appropriate to evaluate the transmission function, that controls   quantum transport  of charge and heat through the system at the atomistic level.  Here we adopt  the linear regime for the difference of the Fermi functions of the left and right reservoirs, $f(E, \mu_L, T_L)- f(E, \mu_R, T_R)$; moreover, 
for  sake of simplicity, we consider pure electronic transport.
In the particular case  that many-body effects (such as electron-electron, electron-phonon or phonon-phonon interactions~\cite{NICOLIC12}) are made negligible, the Landauer approach is recovered~\cite{DATTA03,JEONG10}. 
Anyhow, if many body interactions are present in the central device,
the Keldysh formalism can anyway encompass at the appropriate level of approximation wide classes of many-body scattering processes,  and we here mention just as an example the successful proposal of electron-phonon interaction in the   lowest order approximation~\cite{FRED05,FRED014,FRED16,CUEVAS05}, and other possible  analytic simplifications.\cite{BEVI16}

The key ingredient for the description of transport in the spirit  of the Keldysh formalism and mean field approach, is the
electronic transmission function ${\mathcal T} (E)$ which contains the microscopic physics of the sample under temperature and chemical potential differences, and its connection with the leads.
Numerous first-principle calculations have been proposed based on density functional theory in the Green's function many-body formalism to study electronic  and thermal  conductances in nanoscale and molecular systems~\cite{SANVITO06,NIKO12,TAKAI16,MINGO06,WANG08}, often combined with tight binding Hamiltonians.\cite{NICOLIC12,DATTA03}

To pick-up the essentials of charge  and electronic thermal contribution to coherent transport in TE, in this paper we do not go through ab initio evaluation of the transmission function, but we focus on  special functional shapes, such as  Fano transmission functions and Lorentzian resonances and antiresonances, most frequently encountered in the actual transmission profiles of nanostructured systems, due to quantum interference effects.  
In particular, the  review by Lambert~\cite{LAMB15} on quantum interference effects in single-molecule electronic transport has underlined the importance of recognizing the peak and dip  nature in the evaluated landscape of ${\mathcal T} (E)$ and how they can be tuned by appropriate system parameters, as recently implemented also by stretching.\cite{TORRES15,TSUT15}
For instance, in a molecular system coupled to electrodes, Breit-Wigner (Lorentzian)-like~\cite{BW36} transmission function occurs at electron energies  which approach the energies of the composing orbitals for sufficiently spaced  molecular levels. On the other side, the ubiquitous asymmetric Fano like resonances~\cite{FANO61,MIRO10,Huang15,SSP} may occur e.g.  in chains of molecular systems with attached groups when  the energy of the electron resonates with a bound state of the pendant group~\cite{CHAKRA06,FARCHION12}.

Impact of Breit-Wigner and Fano transmission shapes on the TE properties of nanostructured materials  have been recognized for graphene quantum rings~\cite{ORELLANA15}  and nanoribbons~\cite{NICOLIC12,GONG13} but also for quantum dots~\cite{SANCHEZ15,ORELLANA12}, and in the vast field of molecular electronics~\cite{CUEVAS10, FARC96} for nanoscale molecular bridges and molecular wires,\cite{LAMB14,SUAREZ13,LAMB09,MURPHY08,LAMB06} and molecular constrictions. { Noticeably, molecular junctions have been proposed~\cite{MURPHY08} as optimal candidates for large values of the figure of merit $ZT$.}

Our paper aims to a systematic study of paradigmatic model nanosystems, because of their own interest and in order to infer guidelines for optimal efficiency and thermopower of actual TE quantum structures. To keep the presentation reasonably self-contained, in Section II we summarize relevant aspects of quantum transport for molecular devices, in the linear response regime. In Section III we elaborate the transport parameters with some significant rationalization.  In particular a novel expression  of the efficiency of the device is worked out. Convenient expressions of electric conductance, thermopower coefficient, thermal conductance, power output, Lorenz function, performance parameter and efficiency are reported in terms of kinetic parameters defined in dimensionless form.
In Section IV  and in Section V  the general concepts are  specified  in the case of the Fano transmission function and the encompassed Lorentzian resonances and antiresonances; it is remarkable and rewarding that the treatment becomes fully analytic, in terms of polygamma functions and Bernoulli numbers. 
This permits deeper physical insight on the variegated aspects of carrier transport and the instructive numerical simulations  reported in Section VI. By virtue of our  procedure, analytic in a wide extent and fully analytic in a number of significant limits in the parameter domain, universal features describing transport in Fano-like models emerge with great evidence. This is of major interest on its own right; also, and more importantly, the universal curves may provide useful guidelines for realistic nanosystems, whose transmission lineshapes can be tailored and fitted with the studied models in some appropriate energy ranges. Section VII contains the conclusions.

\section{Transport equations in the  linear response regime for molecular devices}

The transport equations of a nanoscale system of non-interacting 
   electrons are essentially controlled by the transmission 
   function ${\mathcal T} (E)$.  
   The   charge (electric)   current $I_e$,  the  left and the right 
   heat (thermal)  currents $I_Q^{(left)}$ and $I_Q^{(right)}$, the input or  
    output power ${\mathcal P}$ (with ${\mathcal P}>0$ 
    in power-generators, and ${\mathcal P}<0$ in  refrigerators),  the efficiency $\eta$ 
    (in power generation) and the efficiency $\eta_{refr}$ (in refrigeration)
    due to the transport of (spinless) electrons across 
    a mesoscopic device  in stationary conditions are given by the  expressions~\cite{GOOD09,CUEVAS10}
        \begin{subequations}
        \begin{eqnarray}
           & &  I_e= I_e^{(left)} = I_e^{(right)} 
                     = \frac{- e}{h} \int_{-\infty}^{+\infty} dE
                 \, {\mathcal T} (E)  \left[ f_{L}(E) - f_{R}(E) \right]  
                                               \\ [2mm]
         & & I_{Q}^{(left)} = \frac{1}{h} \int_{-\infty}^{+\infty} dE   (E- \mu_L)  
                \,  {\mathcal T}(E)  \left[ f_{L}(E) - f_{R}(E) \right]  
                                                \\ [2mm]
         & & I_{Q}^{(right)} = \frac{1}{h} \int_{-\infty}^{+\infty} dE   (E- \mu_R)  
                \,  {\mathcal T}(E)  \left[ f_{L}(E) - f_{R}(E) \right]  
                                              \\[2mm]
         & &   {\mathcal P} = I_{Q}^{(left)} - I_{Q}^{(right)} 
                          =  \frac{1}{h} \, (\mu_R -\mu_L)  \int  dE
                 \, {\mathcal T} (E)  \left[ f_{L}(E) - f_{R}(E) \right] 
                                                \\[2mm]
          & &   \eta =  \frac{ I_{Q}^{(left)} -  I_{Q}^{(right)} } { I_{Q}^{(left)} } 
                    =   (\mu_R - \mu_L) \frac{ \int dE
                 \, {\mathcal T} (E)  \left[ f_{L}(E) - f_{R}(E) \right]   }
                 { \int dE (E- \mu_L)
                 \, {\mathcal T} (E)  \left[ f_{L}(E) - f_{R}(E) \right]  }   
                                                \\[2mm]
          & & \eta_{refr} = \frac{ I_{Q}^{(right)} } { I_{Q}^{(left)} - I_{Q}^{(right)}} 
                    =   \frac{ \int dE  (E - \mu_R )
                 \, {\mathcal T} (E)  \left[ f_{L}(E) - f_{R}(E) \right]   }
                 { \int dE (\mu_R- \mu_L)
                 \, {\mathcal T} (E)  \left[ f_{L}(E) - f_{R}(E) \right]  }   
            \end{eqnarray} 
            \end{subequations}
     where $e=|e|$ is the absolute value of the electronic charge. The 
     positive direction in the one-dimensional device has been chosen 
     from the left reservoir to the central  device  in the left lead, and from 
     the central device to the right reservoir in the right lead. Notice 
     that Eqs.(1) hold in the linear and non-linear regime, and 
     apply to thermal devices, regardless if they act as output-power 
     generators  or input-power 
      absorbers  (i.e. {\it refrigerators}, often addressed as {\it heat pumps}). 
     
   We are here interested in the 
 linear response  of the system and assume 
   that $\Delta \mu = \mu_L -\mu_R$ and $\Delta T = T_L -T_R$ can be treated as
   infinitesimal quantities.  For power 
     generators, the appropriate  
   operative conditions can be specified as follows:
    
    (i)  { Without loss of generality, from now on, it is  assumed that the 
    left reservoir  is  the hot  one and the right reservoir is the cold one}, 
    namely:
             \begin{subequations} 
             \begin{equation}
                       \Delta T = T_L -T_R  >0 \ .
             \end{equation} 
   The quantity $\Delta T$ is always positive (regardless if finite or 
    infinitesimal); on the contrary,  the  sign of the
    quantity $\Delta \mu$ is controlled or chosen case-by-case.
    
    (ii) The power 
     generators mimic in principle a macroscopic thermal 
    machine  if heat is extracted  from the hot reservoir and a fraction of 
    it is transmitted to the cold  reservoir. This  entails that both the left 
    heat current and the right heat currents  are  positive, and the former 
    is larger than the  latter; namely:
             \begin{equation}
                I_{Q}^{(left)} > I_{Q}^{(right)} >0   .
            \end{equation}   
    The difference of the left and right thermal currents represents the 
    output power of the  nanoscale thermal generator.
     In Fig.1 we report schematically the picture of  transport through 
    nanoscale power generator.

   \begin{figure}[h]
\begin{center}
\includegraphics{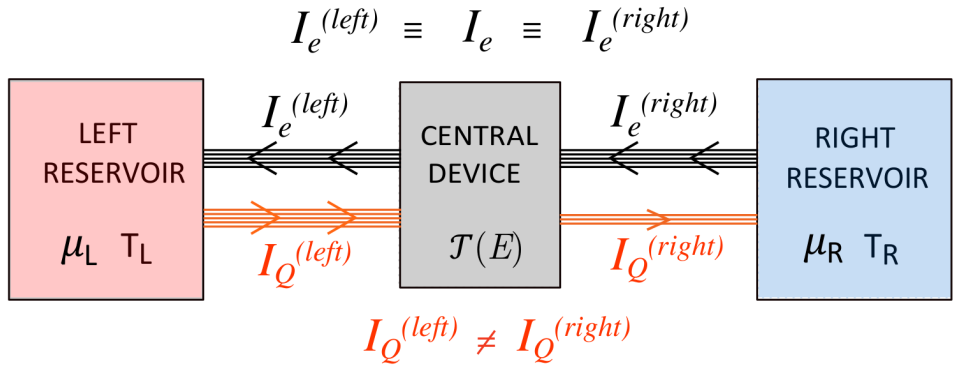}
\end{center}
\vspace{-0.8cm}
\caption{Schematic representation of 
   a two-terminal  power-generator device, with electron transmission 
   function ${\mathcal T}(E)$, and $T_L>T_R$. In steady situation, the 
   charge current  is conserved  in the left and right electrodes.    
   Heat current is not conserved,  and heat flowing from the hot  source 
   is partially transmitted  to the cold one. 
  [For refrigerators, maintaining   $T_L>T_R$,  the arrows must 
  be oriented in the opposite direction].}
\end{figure}
   The situation of refrigerators could be 
    dealt with in a similar way: in the cooling mode the 
    nanoscale device   satisfies the conditions
             \begin{equation}
                I_{Q}^{(left)} < I_{Q}^{(right)} < 0  \ ,
            \end{equation}   
            \end{subequations}
     with heat flowing from the cool reservoir to the hot one; the 
     difference  of the left and right thermal currents represents the power 
    absorbed   from the  nanoscale thermal refrigerator.  In this work
      we consider explicitly only the case of power generation, 
     since the case of power absorption is  akin. 
     
         \vspace{0.5cm}
     \noindent       {\bf Linearization of the transport equations}
     \vspace{0.5cm}          
      
  \noindent  Consider the Fermi  distribution function 
         \begin{equation*}
                f(E;\mu,T)
                       = \frac{1}{ e^{(E-\mu)/ k_BT} + 1}  \equiv f(E)  \ ;
          \end{equation*} 
    the derivatives with 
   respect to the energy, the temperature and the chemical potential are 
   linked   by the relations
       \begin{equation}
          \frac{ \partial f}{\partial  E}= - \frac{1}{k_BT} \cdot
                   \frac{  e^{(E-\mu)/ k_BT} }
                    { [e^{(E-\mu)/ k_BT} + 1]^2} 
                           \quad ; \quad 
                               \frac{ \partial f}{\partial  \mu}  
                     \equiv  -  \frac{ \partial f}{\partial  E}   
                          \quad ; \quad         \frac{ \partial f}{\partial T}   
                 \equiv  \frac{E-\mu}{T} \, ( -  \frac{ \partial f}{\partial  E} ) \ .
      \end{equation}  
    
   In the linear approximation,  the  Fermi function of the right reservoir  
   can be expanded in terms of the Fermi function of the left 
   reservoir in the form
        \begin{equation*}
             f(E;\mu_R,T_R) = f(E;\mu_L,T_L) 
                                   +(\mu_R -\mu_L) \frac{\partial f_L}{\partial \mu_L}
                                   + (T_R -T_L) \frac{\partial f_L}{\partial T_L} \ .
       \end{equation*}
   We denote by $\Delta T$ (with $\Delta T>0$) the  temperature  difference between
    the left and right reservoir, with $\Delta \mu$ the  difference of the chemical
   potential, and with $\Delta V$ the applied bias; namely
          \begin{equation}   
                 \Delta T  = T_{L} - T_{R} \ (>0) \ \ , \ \  
                \Delta \mu = \mu_{L} - \mu_{R} \ \ , \ \ 
                \Delta \mu = (-e) \Delta V \ \ , \ \
                \Delta V = V_{L} - V_{R} \ .  
           \end{equation}  
    It follows
            \begin{equation}  
                f_{L}(E) - f_{R}(E) = ( -  \frac{ \partial f_L}{\partial  E} )
                   \left[  \Delta \mu  \ + \  (E-\mu_L) \, \frac{ \Delta T}{T_L} \right]  .
           \end{equation} 
           
   The transport equations (1) for charge current,  heat current,   
   power-output  and the efficiency parameter become  
   for low voltage bias and low temperature bias:
   
        \begin{subequations}
        \begin{eqnarray}
             I_e &=& \frac{-e}{h} \int dE
                      \, {\mathcal T} (E) \, ( -  \frac{ \partial f_L}{ \partial  E} )
                   \left[ -e \,\Delta V + (E-\mu_L) \dfrac{ \Delta T}{T_L} \right]  
                                            \\ [3mm] 
             I_{Q}^{(left)} &=& \frac{1}{h} \int dE \,   (E- \mu_L)  
                \,  {\mathcal T}(E) \, ( -  \frac{ \partial f_L}{ \partial  E} )
                    \left[ -e \,\Delta V + (E-\mu_L) \frac{ \Delta T}{T_L} \right]
                                         \\[3mm] 
                 {\mathcal P} &=&   \frac{1}{h} \, e \, \Delta V \int dE
                      \, {\mathcal T} (E) \, ( -  \frac{ \partial f_L}{ \partial  E} )
                        \left[ -e \,\Delta V + (E-\mu_L) \frac{ \Delta T}{T_L} \right] 
                                            \\[2mm] 
                  \eta &=&   e \, \Delta V  \frac{ \int dE
                      \, {\mathcal T} (E) \, ( -  \dfrac{ \partial f_L}{ \partial  E} )
                        \left[ -e \,\Delta V + (E-\mu_L) \dfrac{ \Delta T}{T_L} \right] }
                        {  \int dE \,   (E- \mu_L)  
                \,  {\mathcal T}(E) \, ( -  \dfrac{ \partial f_L}{ \partial  E} )
                    \left[ -e \,\Delta V + (E-\mu_L) \dfrac{ \Delta T}{T_L} \right] }  \,.
            \end{eqnarray}  
           \end{subequations}

       At this stage, in the conventional elaboration of the transport properties  
       of nanoscale  systems, it is customary to introduce the kinetic transport
      coefficients $L_0,L_1,L_2$   usually in the form
             \begin{equation*} 
              L_n = \frac{1}{h} \int dE   \, {\mathcal T} (E) (E-\mu_L)^n 
                             ( -  \frac{ \partial f_L}{\partial  E} ) \  \qquad (n=0,1,2) \ . 
             \end{equation*} 
       It is seen by inspection that the electric  charge current  and the heat 
       current  $(I_e, I_Q^{(left)})$ are linked to the bias potential and 
       bias  temperature   $(\Delta V, \Delta T)$ via a $2\times 2$ matrix, 
       controlled  by $L_{1,2,3}$. It is also apparent  that the units 
       of  the coefficients $L_n$ change with $n$ and are given 
      by $\ {\rm (eV)}^{n-1} \cdot {\rm sec^{-1}} \ $.

      For the purpose of this article, that focuses on   performance of 
      devices, optimization conditions, and 
      comparison   of  transmission functions, 
       it is useful (and ``practically necessary") 
      to clearly disentangle  quantities under elaboration from the 
      entailed units of measure. For a deeper understanding of the 
      physics of transport  processes, and also for computational purposes,
      it is preferable  and rewarding to  process  dimensionless quantities,
      adopting units based on fundamental constants 
      or combination of fundamental constants, as shown in detail 
      in the next section.


       \section{Dimensionless kinetic parameters  and natural units 
         for   nanoscale devices}

     The structure of Eqs.(6), and the previous discussed motivations,  
     suggest  to define the {\it dimensionless kinetic transport 
     coefficients} $K_n$    as follows
        \begin{equation} 
             K_n =  \int dE   \, {\mathcal T} (E) \, \frac{(E-\mu)^n}{(k_BT)^n}
                                     \, ( -  \frac{ \partial f}{\partial  E} ) 
                                    = K_n(\mu,T)  \quad (n=0,1,2) \ ,
       \end{equation}  
   where $\mu= \mu_L \ , \ T=T_L \ {\rm and} \ f= f_L$. It is  apparent 
   that   $K_0$ and $K_2$ are positive quantities,  while $K_1$ can be 
   either positive or negative; furthermore $K_1$ certainly vanishes 
   whenever  the transmission function is  an even function with respect 
   to the chemical potential.
   
  The expression of the kinetic coefficients $K_n$ can be conveniently
   worked out with the Sommerfeld expansion~\cite{SSP}, provided the 
  transmission function is reasonably smooth on the scale of 
  the thermal energy $k_BT$ (which is the energy scale of the 
  derivative of the Fermi function).  In the treatment of nanostructures the
  Sommerfeld expansion is hardly applicable,  and other 
  procedures must be considered. In the paradigmatic case  of Fano 
  transmission function and alike,  we show in Appendix A that 
   the kinetic   transport coefficients can be obtained analytically.

  From the structure of Eq.(7), it can be noticed that the 
  expressions $K_0,K_1,K_2$ are the zero, first 
  and second moment  of the definite positive function, given by the 
  product the transmission function times  the opposite of the derivative 
  of the Fermi function. The moments  of any definite  positive function 
  satisfy  basic and general restrictions, and in particular 
  for $K_{0,1,2}$ it holds
           \begin{equation}
             \frac{K_2}{K_0}  \ge \left(  \frac{K_1}{K_0} \right)^2
                               \quad \Longleftrightarrow \quad
                       K_2  \, \ge \, \frac{K_1^2}{K_0}    
                      \quad \Longleftrightarrow \quad 
                        \frac{K_1^2}{K_0K_2} \le 1  \ . 
         \end{equation} 
   We exploit the above inequality for defining  a  novel key  parameter 
   of far reaching significance
            \begin{equation}
                     p= \frac{K_1^2}{K_0K_2}     \qquad({\rm with} \ \ 0\le p \le 1) \ .
            \end{equation}  
   The so defined  {\it  p-performance parameter}  is dimensionless and confined in 
   the interval from  zero to unity.  The upper bound holds only  when the 
   energy spread of the 
  definite positive integrand in Eq.(7)  vanishes.  The lower bound holds  
  when  $K_1=0$, and in particular whenever the transmission 
  function is even with respect to the chemical potential.

   The  performance parameter  $p$  
   characterizes and controls 
   the efficiency  of the nanoscale thermal device, as we show in detail in Appendix B. 
   It is remarkable that the optimal efficiency $\eta$ of the device, inferred
   from Eq.(6d), is linked   to the $p$-performance parameter by the simple 
   expression:
                \begin{equation} 
                  \frac{\eta}{\eta_c} =  \frac{  1 - \sqrt{ 1 - p}  }
                                        {  1 + \sqrt{ 1 - p}  }   \ ,
             \end{equation} 
      where
              \begin{equation}
                        \eta_c \equiv \frac{\Delta T}{T} \equiv \frac{T_L - T_R}{T_L}
                              \quad (T=T_L > T_R) \ 
               \end{equation} 
     is the efficiency of the ideal Carnot cycle.  It is almost superfluous to 
     add that the optimal efficiency  of the  device, provided by Eq.(10)  
     is smaller than   the Carnot cycle efficiency,  as required 
     by the general principles of thermodynamics.  
       It is also apparent that  the 
      efficiency $\eta$ takes its maximum  value $\eta_c$ 
      for $p=1$, and decreases monotonically to zero for decreasing 
      values of  $p$.

      We now insert into Eqs.(6) the kinetic transport parameters   
      defined in Eqs.(7). To simplify a little bit  the 
   notations (with attention to avoid ambiguities), in  Eqs.(7) the  
   temperature  $T_L$  and the chemical potential $\mu_L$ for the left 
   reservoir are denoted dropping the subscript $L$ for left, 
   i.e.  $T_L \rightarrow T$ and $\mu_L \rightarrow \mu$; the same 
   simplified notation is applied to Eqs.(6). Then, the transport 
  equations (6)  take  the  compact and  significant  form
            \begin{subequations}
            \begin{eqnarray}
                I_e &=&  \frac{e^2}{h}  K_0 \, \Delta V 
              - \frac{e^2}{h} \frac{k_BT}{e}  K_1   \frac{\Delta T}{T}
                                       \\[2mm] 
                I_Q^{(left)}  &=& - \frac{e^2}{h} \frac{k_BT}{e}  
                     K_1\,  \Delta V +  \frac{e^2}{h} \frac{k_B^2T^2}{e^2}  
                                         K_2  \frac{\Delta T}{T}
                                                 \\[2mm] 
                {\mathcal P} &=&   - \frac{e^2}{h}  K_0 \, (\Delta V)^2  
                         + \frac{e^2}{h} \frac{k_BT}{e} K_1 
                          \, \Delta V \, \frac{\Delta T}{T}
                                   \ \left[ \equiv  I_Q^{(left)} -  I_Q^{(right)} \right]
                                              \\[2mm] 
                  \eta &=&  \frac{ \ \ - K_0 \, (\Delta V)^2  
      \ + \ \dfrac{k_BT}{e}  K_1 \, \Delta V \, \dfrac{\Delta T}{ T} }
                         {  - \dfrac{k_BT}{e}  K_1\,  \Delta V 
                         + \dfrac{k_B^2 T^2}{ e^2} K_2 \, \dfrac{\Delta T }{ T} } 
                        \  \left[\equiv \frac{ {\mathcal P} }{ I_Q^{(left)} } \right] \ .
         \end{eqnarray} 
         \end{subequations}
   The ingredients of Eqs.(12) involve the dimensionless kinetic 
   parameters $K_{0,1,2}$, and the  Carnot 
   efficiency $\eta_c$
    of an ideal 
   device working  between the temperatures $T_L >T_R$. 
   Eqs.(12) also contain the  applied bias potential  $\Delta V$, and the 
   so called  ``thermal potential" $\phi_T$,  defined by the 
   relation $ \phi_T \equiv k_B T/e$.  The quantum of conductance $e^2/h$ 
   also appears naturally.

    Using Eqs.(12), the transport  coefficients of  interest 
   in measurements,  such as the electric conductance, the Seebeck
   coefficient,  the thermal conductance, the Lorenz number, the power 
   output and the efficiency parameter,  can be worked  out as follows.

    Consider  first the  thermoelectric system in {\it isothermal situation},
    i.e. with the electrodes kept at the same temperature.   
    Eq.(12a) in the absence of temperature gradients gives
        \begin{equation}
            \Delta T \equiv 0 \ \ \Longrightarrow \ \ 
              I_e =  \frac{e^2}{h} K_0  \, \Delta V \equiv  \sigma_0 \, \Delta V   
               \quad {\rm with} \quad 
                \ \sigma_0   =  K_0 \, \frac{e^2}{h} \ .
       \end{equation} 
    The isothermal conductance $\sigma_0 $ represents the
    proportionality coefficient between the  electric current and the applied
    voltage $\Delta V$, with no temperature gradient across the sample.

   In the  general situation when a voltage and a temperature gradient 
   are both applied  to the thermoelectric system, the electric current  
   given by  Eq.(12a) can be written in the more effective form 
         \begin{eqnarray}
              I_e &=&  \frac{e^2}{h} \, K_0  \left[  \Delta V  
                -  \frac{K_1}{K_0} \, \frac{k_BT}{e} \, \frac{\Delta T}{T} \right]
                                 \nonumber\\[2mm]
              &=& \sigma_0 \left[ \, \Delta V  + S \, \Delta T \right]
                \quad  {\rm with} \quad  
                     S(T,\mu) = -  \frac{K_1}{K_0} \, \frac{k_B}{e}  \ ;
         \end{eqnarray} 
     the contribution to the electric current, proportional to  the temperature
    bias,   defines the   thermoelectric power or  Seebeck coefficient $S$.
    In open circuit situation, we have $I_e=0$; this means that  
    the thermoelectric power represents essentially the  potential drop 
    for unitary temperature gradient  for zero electric current.

     From  Eq.(12a) we can extract for  $\Delta V$ the expression:
              \begin{equation*}
             \Delta V =  \frac{1}{(e^2/h) K_0}  I_e  
                     +  \frac{ K_1}{K_0} \, \frac{k_BT}{e} \, \frac{\Delta T}{T} \ .
          \end{equation*}
    Replacement of  such a value  into   Eq.(12b) gives 
          \begin{equation*}
               I_Q^{(left)} = -\frac{e^2}{h} \, \frac{k_BT}{e} \, K_1 
                    \left[  \frac{1}{ (e^2/h) K_0}  I_e  
                     +  \frac{  K_1}{K_0} \, \frac{k_BT}{e} \, \frac{\Delta T}{T} \right]   
                  +  \frac{e^2}{h} \, \frac{k_B^2T^2}{e} \,  K_2 \,  \frac{\Delta T}{T}  \ .
          \end{equation*}
      Then
               \begin{equation}
             I_Q^{(left)}  = - \frac{K_1}{K_0} \, \frac{k_BT}{e} \, I_e + \kappa_{el} \Delta T  
                        \qquad {\rm with} \qquad
                   \kappa_{el} = T  \, ( K_2 - \frac{K_1^2}{K_0} ) \, \frac{k_B^2}{h} \ ,
          \end{equation} 
   where $\kappa_{el}$ defines the  electronic contribution to the thermal conductance of the system (heat 
   current per unit temperature gradient for zero electric current).
   The ratio between the thermal conductance and the electric 
   conductance is called the Lorenz number; it is given by 
      \begin{subequations}
         \begin{equation} 
             L = \frac{\kappa_{el}}{\sigma_0 T} = 
                             \frac{K_0 K_2 - K_1^2} {K_0^2} \, \frac{k_B^2}{e^2}  \ .
        \end{equation}   
   Thermal conductance and  Lorenz number are essentially positive quantities,
    as can be inferred  from their physical meaning and from  the inequality (8). 
 { Another  parameter traditionally used in the literature is the dimensionless  {\it figure of merit}. Neglecting lattice conductance, the figure of merit  for electrons carrier transport reads
     \begin{equation} 
      (ZT)_{el}=\frac{T\sigma_0S^2}{\kappa_{el}}=\frac{S^2}{L}=\frac{K_1^2}{K_0 K_2 - K_1^2}  \ .
       \end{equation}   
  From Eq.(9) and Eq.(16b) one can see that the $(ZT)_{el}$ and $p$ parameters  are linked by the relations
   \begin{equation} 
   (ZT)_{el}=\frac{p}{1-p}  \quad  \Longrightarrow \quad  p=\frac{(ZT)_{el}}{(ZT)_{el} +1}
   \end{equation}  }
     \end{subequations}  
     
 \vspace{0.2cm}   
\noindent {\bf The operative conditions for molecular power-generators }
\vspace{0.2cm}

     The operative conditions for molecular power-generators imply a positive 
     power-output;  such requirement  using Eq.(12c) reads
         \begin{equation}
               {\mathcal P} = - \frac{e^2}{h} K_0 \, (\Delta V)^2   
                + \frac{e^2}{h} \, \frac{k_BT}{e} \, K_1
                        \, \Delta V \, \frac{\Delta T }{T}     > 0 \ .
         \end{equation} 
    Since $\Delta T$ and $K_0$ are both positive, a necessary condition 
    to satisfy Eq.(17) is that $K_1$ and $\Delta V$ have the same sign. 
    The output power vanishes for 
           \begin{equation*}
               \Delta V =0 \quad {\rm and} \quad
                 \Delta V =  \frac{K_1}{K_0} \, \frac{k_BT}{e}  \, \frac{\Delta T}{T} \ .
            \end{equation*}
    Suppose we have chosen the parameters $T, \mu$ for the left 
    reservoir (the hotter of the two reservoirs), and also fix  $\Delta T (>0)$.
    The only variable parameter in Eq.(17) remains $\Delta V$. 
     It is apparent that 
            \begin{subequations}
             \begin{eqnarray}
               {\rm if} \quad K_1>0 \quad \Longrightarrow \quad 
                 {\mathcal P}> 0  & \quad {\rm for} \quad &    
                 0 < \Delta V <   \frac{K_1}{K_0} \, \frac{k_BT}{e} \, \frac{\Delta T}{T} 
                                              \\[2mm] 
               {\rm  if} \quad K_1<0 \quad  \Longrightarrow \quad
                {\mathcal P}> 0   & \quad {\rm for} \quad & 
                 \frac{K_1}{K_0} \, \frac{k_BT}{e} \, \frac{\Delta T}{T} < \Delta V < 0 \  . 
             \end{eqnarray}  
             \end{subequations}
    The optimized maximum value of the power-output occurs at midway of 
    the intervals  indicated in Eqs.(18), and reads
              \begin{equation}
                    {\mathcal P} =  \frac{1}{4}  \, 
                      \frac{K_1^2}{K_0} \,\frac{k_B^2T^2}{h} \, (\frac{\Delta T}{T})^2
                   = \frac{1}{4} \frac{K_1^2}{K_0} \,\frac{k_B^2}{h} \, T^2 \, \eta_c^2 \ .             
              \end{equation} 
              
\

\

 \noindent {\bf Natural units  for   nanoscale devices}
   \vspace{0.2cm}

      For sake of completeness we briefly  summarize the natural units 
      encountered so far.
      The natural unit of conductance is given by the quantum 
      of conductance 
             \begin{subequations} 
             \begin{equation}  
                 \frac{e^2}{h}  = \frac{1}{25812.807} \, \Omega^{-1}
                       = 3.874046 \cdot 10^{-5} \  \frac{\rm A}{\rm V} \ ;
              \end{equation} 
      the value is based on the von Klitzing constant $h/e^2$, 
      whose  experimental accuracy is better than eight significant digits.
      The conductance of a single periodic chain in the allowed 
      energy region equals $e^2/h$. 
      
       The natural unit of Seebeck thermoelectric power is
              \begin{equation} 
                    \frac{k_B}{e}  =   86.17 \  \frac{ \mu {\rm V} }{\rm K}  \ .
               \end{equation} 
     Good thermoelectric materials have thermoelectric powers 
     of the order  of $k_B/e$. Notice that the ratio between the Boltzmann 
     constant and the electron charge can also be conveniently replaced 
     by $\phi_T/T$, where $\phi_T = k_BT/e$ is the thermal 
     voltage (a quantity and a concept  embedded in the architecture  
     of electronic circuits  see Ref.\onlinecite{SEDRA10}).
         
    For instance, at room temperature $T_0= 300 \ K$, $\phi_0 \approx 0.025 \ $V,
     and $\phi_0/T_0$ recovers Eq.(20b), as expected.

   For the Lorenz number (or better: for the Lorenz function) the natural 
     unit is given by the square of Eq.(20b); namely
               \begin{equation} 
                  \frac{k_B^2}{e^2}  
                           = 74.25 \cdot 10^{-10} \, \frac{ {\rm V}^2}{ {\rm K}^2} \ .
                \end{equation} 
     And finally for the thermal conductance a useful unit is given by the 
     following combination of universal constants
               \begin{equation} 
                  \frac{k_B^2}{h}  = 1.8 \cdot 10^{6} \ 
                                      \frac{\rm eV}{\rm sec} \cdot \frac{1}{\rm K^2} \ ,
                \end{equation} 
         \end{subequations}
      which can be seen as the counterpart  of Eq.(20a) for the electric 
      conductance.
    
     For convenience,  the  thermoelectric transport parameters, 
     expressed  in terms  of  dimensionless kinetic coefficients and 
     natural units, are summarized  in Table 1.
     \vspace{0.5 cm}

    \begin{table} [t]
  \centering 
  {\begin{tabular}{@{\extracolsep{\fill}}lcr@{}} \vspace{2mm}\\ \hline
  \multicolumn{3}{l}{
  Nanostructure: ${\mathcal T}(E)$  transmission function  }\\
   \multicolumn{3}{l}{
  Dimensionless kinetic parameters: $  K_n =  \int dE  \, {\mathcal T} (E) \dfrac{(E-\mu)^n}{(k_BT)^n} 
                         ( -  \dfrac{ \partial f}{\partial  E} ) $ }
                        \hspace{1cm} \vspace{2mm}\\
 \hline \\ \vspace{3mm}
     $\sigma_0 =   K_0 \,  \dfrac{e^2}{h} $ & \hspace{1cm} $S = -   \dfrac{K_1}{K_0} \, \dfrac{k_B}{e}$ & $  \dfrac{ {\mathcal P} }{\eta_c^2 }= \dfrac{1}{4} \, T^2 \, \dfrac{K_1^2}{K_0}  \, \dfrac{ k_B^2}{h}  $ \\   \vspace{2mm}
   $\kappa_{el} = T ( K_2 - \dfrac{K_1^2}{K_0} )   \,   \dfrac{ k_B^2}{h} $ &    & $L = \dfrac{ \ K_0 K_2 - K_1^2 \ } {K_0^2} \  \dfrac{ k_B^2}{e^2}$ \\ \vspace{2mm}
  $p =  \dfrac{K_1^2}{K_0K_2}  \quad  (0\le p \le 1)$ & \hspace{1cm} $  (ZT)_{el}=\dfrac{p}{1-p}$    & $\dfrac{\eta}{\eta_c} =     \dfrac{1-  \sqrt{1-p} } { 1+ \sqrt{ 1 - p}}$ \vspace{2mm}\\ 
    \hline
  \end{tabular}}
  \caption{ {\footnotesize Transport parameters  in the linear approximation  
       for  thermoelectric materials, with  electronic transmission 
       function ${\mathcal T}(E)$.  The kinetic parameters $K_{0,1,2}$ are 
      defined   in dimensionless form.    The  electric conductance $\sigma_0$, 
      Seebeck coefficient $S$, power-output $ {\mathcal P}$,  thermal conductance $\kappa_{el}$, 
     Lorenz number $L$, performance parameter $p$, figure of merit  $(ZT)_{el} $ and efficiency ${\eta}$ 
     are reported. The quantity  $\eta_c$ denotes 
     the Carnot cycle efficiency $\eta_c  = \Delta T/T$, where $\Delta T$ is 
     the temperature difference between the hot reservoir and 
     the cool one.  } }
  {}
  \end{table}


         \section{Kinetic parameters  for Fano lineshapes in the linear 
                          response regime}

      The Fano lineshape transmission function can be written in the form
           \begin{equation} 
                {\mathcal T}_F(E) = \frac{(E - E_d + q\Gamma_d)^2}
                                                {(E-E_d)^2 + \Gamma_d^2}  \ ,
           \end{equation} 
     where $E_d$ is the intrinsic level of the model, $\Gamma_d (>0)$ 
     is the broadening parameter, and the dimensionless 
     parameter $q$ (supposed real and positive) is the asymmetry profile.

      The  dimensionless kinetic parameters  corresponding to 
      the Fano  transmission
      function can be evaluated analytically  for any  range of the 
      thermal energy.  The kinetic   integrals for the Fano
     transmission probability become
           \begin{eqnarray*} 
                                  K_n=\int_{-\infty}^{+\infty}    dE       
                   \,   \frac{   (E -E_d + q   \Gamma_d)^2 /(k_BT)^2 }
	                        { (E - E_d )^2 /(k_BT)^2  + \Gamma_d^2/(k_BT)^2 } 
	            \, \frac{(E-\mu)^n}{(k_BT)^n} \,  \frac{1}{k_BT}
                   \frac{  e^{  (E-\mu)/k_BT} }{ [e^{(E-\mu)/k_BT} + 1]^2} \ .
         \end{eqnarray*}      
    As usual, it is convenient to introduce the
    dimensionless variables
           \begin{eqnarray*}
           & &  z = \frac{E-\mu}{k_BT}   \quad  ; \quad dz =  \frac{dE}{k_BT} 
                  \quad ; \quad   \gamma= \frac{\Gamma_d}{k_BT} 
                  \quad ; \quad   \varepsilon =  \frac{ E_{d}-\mu}{k_BT}
                                   \\[2mm]
             & &   \frac{E-E_d}{k_BT} =  \frac{(E-\mu) - ( E_d - \mu)}{k_BT}
                        \equiv z - \varepsilon   , 
         \end{eqnarray*}        
     where $\varepsilon$ and $\gamma$ are  two dimensionless 
     parameters that, together with the asymmetry parameter $q$, fully specify 
     the Fano model  under attention. The $\varepsilon$ parameter 
     specifies the  position of   the intrinsic  level $E_d$ relative 
     to the Fermi level  in  units of  thermal energy,  while  $\gamma$   
     specifies the broadening parameter  again in units of thermal energy. 
     The asymmetry  parameter ($q  \approx 1-5$ or so) is 
     often considered as an assigned value of the model, although it is 
     of course a third parameter itself.
     With the indicated substitutions, one obtains
             \begin{eqnarray}
               	            K_n=\int_{-\infty}^{+\infty}   
	            dz   \,  \frac{ (z - \varepsilon + q \gamma)^2 \, z^n } 
                  { (z- \varepsilon)^2  + \gamma^2 } 
	         \, ( -  \frac{ \partial f}{\partial  z} )   \qquad  \ 
	                     {\rm with}  \qquad         f(z) = \frac{1}{e^{z} +1} \ . 
         \end{eqnarray}  
    Notice that for  real arguments $(\varepsilon, \gamma)$  the 
    kinetic coefficients are  real functions, as expected.

    For the calculation of Eq.(22),  it is convenient to elaborate the 
    denominator  using the identity
         \begin{equation*}
           \frac{1}{ (z - \varepsilon)^{2} + \gamma^{2}}      \equiv  
             \frac{i}{2\gamma} \left[  \frac{1} { z -  \varepsilon  + i \gamma}
                       - \frac{1}{  z -  \varepsilon  - i \gamma} \right]  \ .
          \end{equation*}
   The kinetic functions  defined in Eq.(22) can be written  in the form
         \begin{equation*}
             K_n =   \frac{i}{2 \gamma}    \int_{-\infty}^{+\infty} \, dz \,   \left[ 
                   \frac{ (z - \varepsilon + q \gamma)^2 z^n } 
                   {z  - \varepsilon + i \gamma}
                 - \frac{ (z - \varepsilon + q \gamma)^2 z^n}
                  {z-   \varepsilon - i \gamma} \right]
                       \, (- \frac{\partial f}{\partial z})  \ .
          \end{equation*}    
    Taking into account that the parameters $(\varepsilon, \gamma, q)$ 
    are real quantities,  we have
             \begin{equation*}
                  K_n =  2 \, {\rm Re} \left\{ \frac{i}{2 \gamma} 
                 \int_{-\infty}^{+\infty} \, dz \,   
                   \frac{ (z - \varepsilon + q \gamma)^2 z^n} 
                                        {z-  \varepsilon + i \gamma}
                       \, (- \frac{\partial f}{\partial z}) \right \}   .
            \end{equation*} 
     We can thus write  for the kinetic parameters  of the Fano lineshape 
     the expression
             \begin{equation} 
                K_n  =  \frac{1}{\gamma}
               \, {\rm Re} \left\{    i\,  \int_{-\infty}^{+\infty}  dz \,   
                   \frac{ \ z^{n+2}  -  2(\varepsilon  -  q \gamma) z^{n+1} 
                     + (\varepsilon - q \gamma)^2  z^n \ } 
                                  {z-  \varepsilon  + i \gamma}
                       \, (- \frac{\partial f}{\partial z}) \right \}  \ .                                                   
          \end{equation} 
      In Appendix A, we  show that the above integrals can be calculated 
      analytically  by means of the trigamma function, and Bernoulli numbers.
          
      For this purpose, we resort  to the the set of 
      auxiliary complex  functions defined in Eq.(A2), and here repeated  
            \begin{equation}
                    I_n(w) =  i   \int_{-\infty}^{+\infty}  dz \,  
                           \frac{z^n}{ z- w } \, (- \frac{\partial f}{\partial z}) 
                           \qquad {\rm Im} \, w < 0  
            \end{equation} 
    where $w= \varepsilon - i \gamma$ is a complex variable,
   independent from the asymmetry parameter of the Fano lineshape.
    In Appendix A  we  show  that all $I_n$ can be expressed in terms 
   of $I_0$, and   furthermore $I_0$ can be calculated analytically 
   with the trigamma function $\Psi_t$.\cite{ABRA}
   The  kinetic parameters $K_n$ of Eq.(23) can be 
   expressed 
   in the form
           \begin{equation} 
               K_n =  \frac{1}{\gamma}   \, {\rm Re} \left[
                 I_{n+2}(w) - 2(\varepsilon - q \gamma) I_{n+1}(w) 
                         + (\varepsilon -  q \gamma)^2  I_n(w) \right]   \ ,                                                   
          \end{equation}  
          where   
          \begin{equation*}
         I_{0}(w) =   \dfrac{1}{2\pi} \,  \Psi_t(\dfrac{1}{2} + \dfrac{i  w}{2\pi}),   \qquad 
         I_n(w) = i  b_{n-1} + w \, I_{n-1}(w)\,\,\,\, {\rm for}  \,\,\,\, n\ge1, 
            \end{equation*}
           \noindent and  $b_n$ are the Bernoulli-like numbers:
             \begin{equation*}
                 b_0 = 1 \ , \ b_1 = 0 \ , \ b_2 = \frac{\pi^2}{3} \ , \ b_3 = 0 \ ,
                          \   b_4 = \frac{7 \pi^4}{15} \ , \  b_5 = 0 , \, \ b_6 = \frac{31\pi^6}{21} \ ,\ \ldots
             \end{equation*} 
          
   \noindent Then the thermoelectric parameters can be calculated using the expressions 
  summarized in Table 1.
  
  A particular case of the Fano transmission function occurs when the 
  asymmetry parameter vanishes. 
  The antiresonance lineshape, setting $q=0$ into Eq.(21), reads  
         \begin{equation} 
                {\mathcal T}_A(E) = \frac{(E - E_d)^2}
                                                {(E-E_d)^2 + \Gamma_d^2}  \ .
           \end{equation} 
   The kinetic parameters  for the antiresonance lineshape, 
   setting $q=0$ into Eq.(25),  and straigt algebraic elaborations become
      \begin{subequations}
      \begin{eqnarray} 
      K_{0} &=&  1- \frac{\gamma}{2\pi}    {\rm Re} \,\Psi_t(\frac12 + \frac{i  w}{2\pi}), \\[2mm] 
      K_{1}&= &- \frac{\gamma}{2\pi}  {\rm Re}  \left[ w \, \Psi_t(\frac12 + \frac{i w}{2\pi}) \right] ,  \\[2mm]
      K_2 &=& \frac{\pi^2}{3}\!-\gamma^2  -  \frac{\gamma}{2\pi}  {\rm Re} \left[ w^{2} \,  \Psi_t(\frac{1}{2} + \frac{i w}{2\pi}) \, \right].     
      \end{eqnarray}                   
            \end{subequations}

  \section{Kinetic parameters  for Breit-Wigner (Lorentzian) lineshapes in the linear 
      response regime}

      The Lorentzian-like  transmission  lineshape can be written in the form
           \begin{equation} 
                {\mathcal T}_L(E) = \frac{\Gamma_d^2}
                                                {(E-E_d)^2 + \Gamma_d^2}  \ ,
           \end{equation} 
     where $E_d$ is the intrinsic resonance level of the model, 
     and $\Gamma_d (>0)$
     is the  broadening parameter.  The kinetic parameters   corresponding 
     to the Lorentzian  transmission function can be evaluated analytically  
     for any  range of the thermal  energy, chemical potential,  location 
     and broadening of the resonant level.   The hybridization 
     energy $\Gamma_d$ sets the lifetime $\tau= \hbar/\Gamma_d$  
     of the electron  in the 
      quantum system.
      We can consider the Lorentz  transmission as the particular case 
      of the Fano lineshape when  the asymmetry 
      parameter  $q\rightarrow \infty$ (and   division by   $q^2$ is performed). 
      From Eq.(25), that provides the kinetic parameters of the Fano lineshape, 
      we obtain that the kinetic parameter  of the Lorentzian lineshape read
                       \begin{equation}
                           K_n = \gamma \, {\rm Re} \, I_n(w)
                      \end{equation} 
       The explicit values of  $K_0,K_1,K_2$   of interest
      for the treatment of  of thermoelectrics  in the energy windows 
      with Lorentzian transmission function
       are the following:
      
      \noindent $K_{0} \!= \! \frac{\gamma}{2\pi}     {\rm Re}  \Psi_t(\frac12 + \frac{i  w}{2\pi})$, \,\,                           
      $K_{1}\!= \!\frac{\gamma}{2\pi}  {\rm Re}  \left[  w \, \Psi_t(\frac12 + \frac{i w}{2\pi}) \right] $, \,\, 
      $K_2 \!= \!\gamma^2  +  \frac{\gamma}{2\pi}  {\rm Re} \left[ w^{2} \,  \Psi_t(\frac{1}{2} + \frac{i w}{2\pi}) \, \right]$.

   \section{Simulation of model thermoelectrics}  
      
    We consider now some simulations of molecular power-generators, with 
    particular interest to establish  domain regions where the efficiency 
    is as near as possible to unity, and the thermopower is large. 
    We begin with the study of the Lorentzian model  for the transmission function, together with  
    the complementary   case of antiresonance  lineshape. Then we examine the 
    situation of the Fano transmission function. These models, can be 
    solved analytically with the trigamma function and Bernoulli numbers, 
   and provide useful guidelines in the understanding  and designing   thermoelectric
   devices.
      
    
     \vspace{0.7cm} 
   \noindent {\bf A.  Transport through Lorentzian transmission functions}
   \vspace{0.5cm} 
   
    The transport properties through the Lorentzian transmission function 
    are controlled by the two dimensionless 
    parameters $(\varepsilon, \gamma)$: the energy
    parameter $\varepsilon= (E_d-\mu)/k_BT$  specifies the position of the  
    electronic level  of the quantum system with respect to the chemical potential in units 
    of thermal energy; the second one $\gamma= \Gamma/k_BT$ specifies  
    the lineshape broadening again in units of the thermal energy. Small 
    values of $\gamma$ (typically $\gamma < 1$) characterize long
     lifetime electronic  states, while large values of $\gamma$ (typically $\gamma > 1$)
    characterize short  lifetime electronic states.
   
      \begin{figure}[b]
	\begin{center}
   \includegraphics{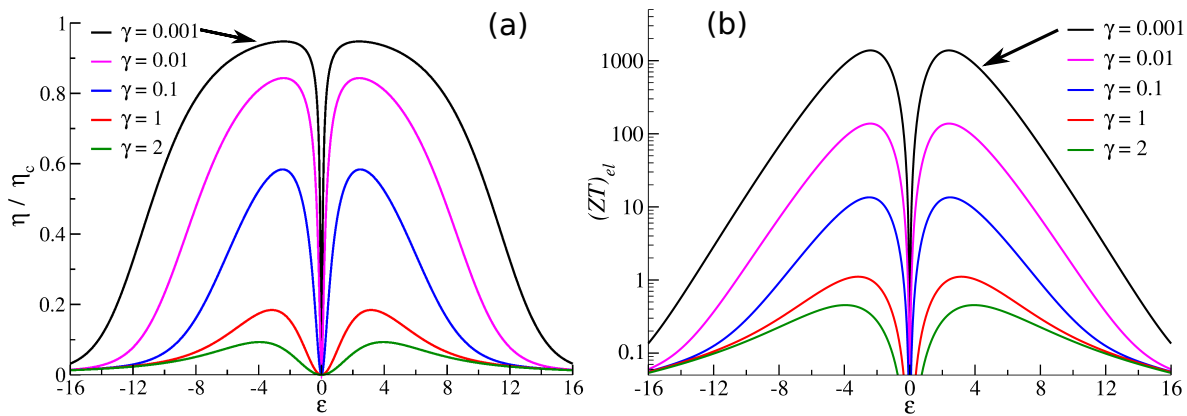}
	\end{center}
	\vspace{-0.8cm}
	\caption{Universal  curves
   for: (a) the efficiency $\eta/\eta_c$, and (b) the figure of merit  $(ZT)_{el}$, of the  thermal machine with Lorentzian 
   lineshape  as a function of the dimensionless energy 
   parameter $\varepsilon= (E_d-\mu)/k_BT$,  for fixed values  of the
    dimensionless broadening  parameter $\gamma= \Gamma/k_BT$. Notice the logarithmic scale on the vertical axis of panel (b). }
\end{figure}

     We begin with the discussion of the behavior of the (relative)  
    efficiency $\eta/\eta_c$, and we report in Fig.2a the family of universal 
   curves   for the efficiency of the  thermal machine with Lorentzian lineshape  transmission 
   as a function of the dimensionless parameter $\varepsilon$,
   for fixed values  of the broadening dimensionless 
   parameter. The values chosen for the broadening parameter 
   are the set of values $\gamma= 2, 1, 0.1, 0.01, 0.001$;  in the case  of a  thermal
   machine operating around room temperature  the set corresponds  
   to the values $\Gamma = 50 , 25, 2.5, 0.25, 0.025$ \ meV.

    It can be noticed  that  the plots in Fig.2a are symmetric   with respect to  $\varepsilon$, 
    and  approach zero  for vanishing $\varepsilon$ and for large $\varepsilon$; 
    this can also be confirmed by 
    appropriate analytic expansion of the trigamma function.  
    
    From Fig.2a it is seen that  the efficiency 
    takes its optimal values for $\varepsilon \approx 2-4$ (or so) for most values 
    of   the broadening parameter. In this range  of  $\varepsilon$ values, 
    the efficiency for long-lived states  $(\gamma \ll 1)$ 
     is near unity, while for short-lived states $(\gamma \gg 1)$ the  
    efficiency   is rather poor. Thus, the good feature of near unity efficiency
    must be matched (and maybe  to
   some extent conflicting)  with  the simultaneous requirement of rather
   small  broadening. This unavoidable link in Lorentzian lineshapes between good efficiency 
   and tendentially small  broadening  is broken by the
    asymmetry parameter of Fano lineshapes,   and 
   represents a major  point of interest of the Fano structures, as we shall see below.
     
       A transport property of primary interest is the Seebeck  thermopower, 
    and we   examine the parameter-region where the (absolute)
    values of the thermopower are reasonably large, i.e. of the order 
    of $k_B/e$  or so.  From the curves reported in Fig.3, it emerges with 
    evidence that  long-lived  quantum {states} ($\gamma <<1$) are the 
    candidates  for   high thermopower. For molecular devices with  
    Lorentzian lineshapes,  it can be noticed that  the thermoelectric 
   power is positive when the chemical potential is larger than the 
   resonance  energy   (i.e. $ \mu> E_d \Longrightarrow \varepsilon <0$); 
   it is zero (by virtue of 
   the symmetry of the lineshape) at the resonance energy;  it is negative 
   when the chemical potential is smaller than the resonance 
   energy (i.e. $ \mu< E_d \Longrightarrow \varepsilon >0$); it goes to 
   zero for large values of $| \varepsilon |$. In principle, 
   the Seebeck coefficient can assume (absolute) values higher or much 
   higher than $k_B/e$, provided $\gamma$  becomes extremely small. 
  Of course, large values  of  Seebeck coefficients are of interest when 
  the efficiency of the thermal machine is also   advantageous, and 
  nanostructures with the desired parameter characteristics  are 
  experimentally achievable. 
   \begin{figure}[h]
	\begin{center}
\includegraphics{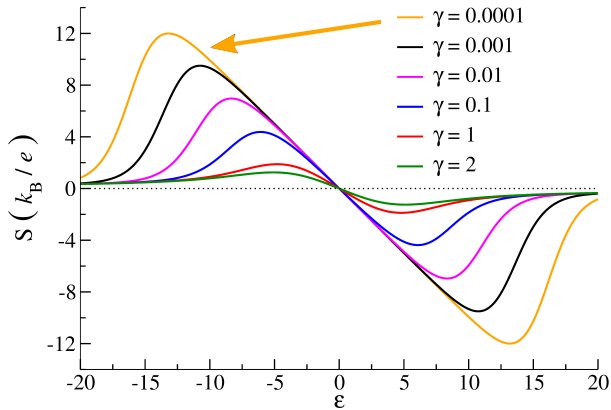}
	\end{center}
		\vspace{-0.8cm}
	\caption{Universal curves for the thermoelectric 
   power (in units $k_B/e$) of  Lorentzian transmission functions, versus  
   the dimensionless energy 
   parameter $\varepsilon= (E_d-\mu)/k_BT$, for fixed values  of the dimensionless broadening  
   parameter $\gamma= \Gamma/k_BT$.  }
\end{figure}    

    \begin{figure}[t]
	\begin{center}
\includegraphics{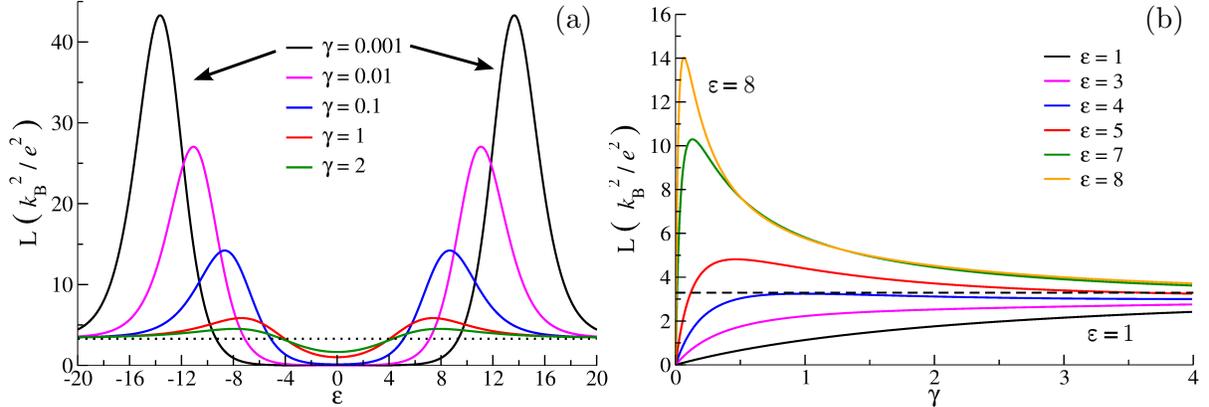}
	\end{center}
		\vspace{-0.8cm}
	\caption{(a) Universal family of  curves
   for the Lorenz number (in units $k_B^2/e^2$) 
   versus the energy parameter $\varepsilon= (E_d-\mu)/k_BT$   for fixed 
   values of  $\gamma= \Gamma/k_BT$  of the resonance transmission function, 
   and (b) versus the broadening 
   parameter $\gamma$  for fixed 
   values of $\varepsilon$.  The straight line  
   drawn  at $\pi^2/3$ represents the common asymptotic value of the 
   family of curves  for large values of  $\gamma$.}
\end{figure}    
     
   We report now in Fig.4 the results for the 
   Lorenz number (or better the Lorenz function).  
    It is well known that the Lorenz number approaches 
   the asymptotic  (Bernoulli-like) value of $\pi^2/3$ whenever  
   the transmission function is rather smooth in the thermal energy 
   scale $k_BT$  (provided no node occurs   in the energy interval under attention); 
   this can be shown with  the Sommerfeld expansion, 
    usually applicable in massive macroscopic thermoelectrics.\cite{SSP} 
     From Fig.4a, it can be seen that the family  ofr curves of the Lorenz number are all depressed with respect to 
     $\pi^2/3$ for $|\varepsilon|$ around the origin; then the curves attain values larger (or much larger) than  $\pi^2/3$  for intermediate values of $|\varepsilon|$, and finally go to the Sommerfeld constant $\pi^2/3$  for high  
     $|\varepsilon|$ values. This down and  up behavior is particularly evident for small values of $\gamma$.
    These features 
    are also corroborated by analytic investigations. From Fig.4b, it can also be noticed 
    that  the curves with $\varepsilon < 4$ (or so) go from zero to the 
    asymptotic value,  in tendentially monotonic way; on the contrary, curves with 
    higher values of  $\varepsilon > 4$ (or so)  overcome 
    the asymptotic value before approaching it  for large $\gamma$. 
   Thus for nanoscale devices the Lorenz  number is very far from being 
   constant, and can be both depressed or enhanced   with respect to the
   Sommerfeld constant.  The region of depression or enhancement 
    is  most interesting for the material performance, because 
    the Wiedemann-Franz law is broken and more flexibility in tailoring 
   thermoelectric properties becomes possible.

 \vspace{0.7cm}
\noindent {\bf B.  Transport through antiresonance transmission functions}
  \vspace{0.5cm}  
  
 \noindent   We consider now transport properties through the  antiresonance  
    transmission  function
         \begin{equation}
            {\mathcal T}_A(E) = \frac{ (E-E_d)^2 }{ (E-E_d)^2 + \Gamma_d^2} \ ,
         \end{equation} 
    and compare with the results obtained in the previous subsection  in 
    the case of resonances.  In Fig.5a we report the efficiency $\eta/\eta_c$ 
    of antiresonance  lineshapes versus the energy parameter $\varepsilon$,  for fixed 
    values  of the broadening   parameter $\gamma$.
    It is  apparent that the efficiency curves 
    are  even  with respect to  $\varepsilon$, and go 
    to zero  for small and large $\varepsilon$; this behavior is confirmed by 
    appropriate analytic manipulations. 
   
      \begin{figure}[h]
	\begin{center}
\includegraphics{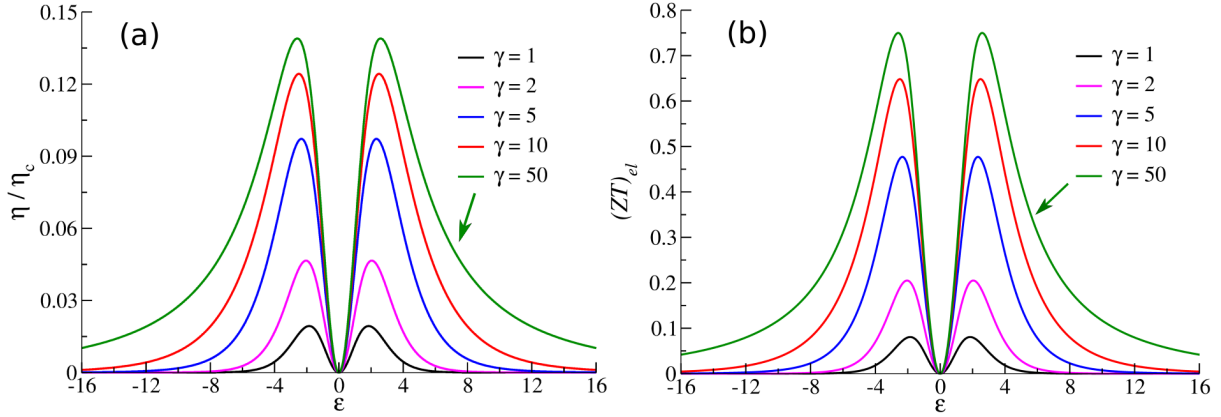}
	\end{center}
		\vspace{-0.8cm}
	\caption{Universal curves for: (a) the efficiency $\eta/\eta_c$,  and (b) the figure of merit  $(ZT)_{el}$, of the 
   thermal machine with  antiresonance lineshape  as a function of the
   energy  parameter $\varepsilon= (E_d-\mu)/k_BT$, for fixed values  of the 
   broadening     parameter $\gamma= \Gamma/k_BT$ . }
\end{figure}    
    
    From Fig.5a, it can also be seen that  the optimal efficiency 
    increases with $\gamma$  up to $\gamma \approx 50$, and then 
    it tends to saturate  (the curve with $\gamma=100 $, not reported, nearly overlaps with the curve with $\gamma=50$). 
    This behavior
   of the 
    efficiency of antiresonances is in striking 
   contrast with the case of Lorentz resonance, where the efficiency 
   always decreases with increasing $\gamma$, as pictured in Fig.2a. 
   Of course the optimal working conditions  of any device are in practice  
   controlled by a trade-off among different requirements,  including 
   efficiency, Seebeck coefficient, actual availability and preparation of  
   materials in the  conditions forecast  as promising
    by the simulations.
        
         In Fig.6 we report the Seebeck  thermopower, for antiresonant levels, 
    characterized  by the $(\varepsilon, \gamma)$ parameters. The  values 
    of the thermopower are  of the order  of $k_B/e$  (or so) for $\gamma$ 
    around unity, and saturate to $\approx 1.8 \, k_B/e$ for larger values of $\gamma$.
     Differently from the behavior of the thermopower of the resonant structure of Fig.3, the Seebeck coefficient of the antiresonance increases with $\gamma$, until it saturates for $\gamma \approx  50$.
    It can be noticed that the thermoelectric power is zero for $\varepsilon =0$ (by virtue of the symmetry of the lineshape), and approaches zero for large $| varepsilon |$; in the region $\varepsilon < 0$ the Seebeck coefficient is negative, while it is positive for $\varepsilon > 0$.
    Thus for the antiresonant structure the Seebeck coefficient is negative for $\mu>E_d$ and is positive for $\mu<E_d$. The opposite signs occur for the resonant structure of Fig.3.  The antiresonant structure produces a Seebeck coefficient with a hole-liken behavior.
   In essence, the comparison 
   of the Seebeck thermopower for  the Breit-Wigner resonance (Fig.3) and for the
   antiresonance (Fig.6) shows  that, in appropriate situations,  it could become 
   preferable to engineer antiresonances, rather than 
   insist in strong peaked structures.
   
       \begin{figure}[t]
	\begin{center}
\includegraphics{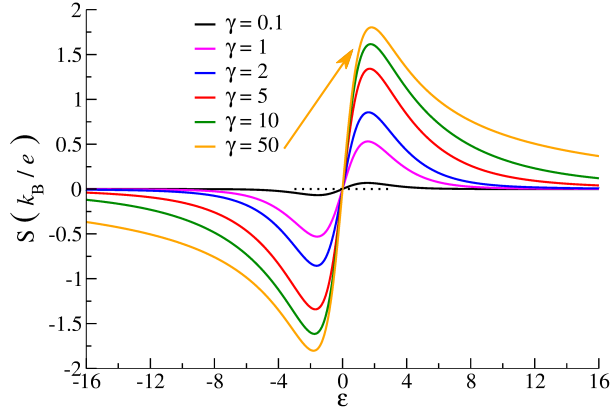}
	\end{center}
		\vspace{-0.8cm}
	\caption{Universal curves of  thermoelectric 
   power (in units $k_B/e$) of antiresonance transmission functions, versus  
   the  energy  parameter $\varepsilon= (E_d-\mu)/k_BT$, for fixed values  of the  broadening  
   parameter $\gamma= \Gamma/k_BT$.}
\end{figure}    

   Fig.7 reports the Lorenz  number as function of the energy parameter $\varepsilon$ for antiresonant structures. The curves reported in Fig.7 are enhanced with respect to the asymptotic value $\pi^2/3$ for  $|\varepsilon|$ around the origin,  present values somewhat smaller than $\pi^2/3$  for intermediate values of $|\varepsilon|$, and finally reach the Sommerfeld constant of $\pi^2/3$ for high values of $|\varepsilon|$. This up and down behavior is particularly evident for high values of $\gamma$. The comparison with the family of curves of Fig.4a for resonant structures, further highlights similarities and differences of resonance and antiresonance structures  in the transmission function.
   
      \begin{figure}[h]
	\begin{center}
 \includegraphics{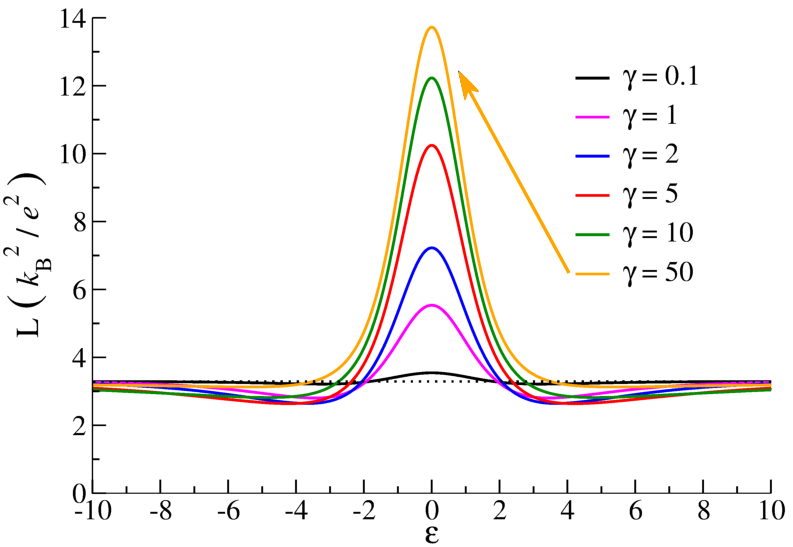}
	\end{center}
		\vspace{-0.8cm}
	\caption{Universal  curves for the  the Lorenz number of antiresonance transmission functions, versus  
   the  energy  parameter $\varepsilon= (E_d-\mu)/k_BT$, for fixed values  of the  broadening  
   parameter $\gamma= \Gamma/k_BT$.}
\end{figure}    
  
    \vspace{0.5cm}
\noindent {\bf  C.  Transport through Fano transmission functions}
   \vspace{0.5cm}
   
\noindent   At this stage  we pass to consider  the  Fano-like lineshapes in the transmission function, which can present in dependence of the intrinsic asymmetry parameter $q$, either a Lorentz 
  resonant level, or antiresonance, or any intermediate structure.
                \begin{equation}
                {\mathcal T}_F(E) = \frac{ (E-E_d + q \Gamma_d)^2 }
                                      { (E-E_d)^2 + \Gamma_d^2} \ .
         \end{equation} 
    The transport properties through the Fano transmission function 
    are controlled by two dimensionless 
    parameters $(\varepsilon, \gamma)$, and by the asymmetry 
    parameter $q$ (assumed to be a  positive number; for negative number 
    the curves must be  reversed); the values $q=0$ and $q= \infty$  
    correspond to the symmetric antiresonance  and symmetric resonance
    Lorentzians,  respectively,  while intermediate values of $q$ 
    produce asymmetric situations. 
    
      \begin{figure}[b]
	\begin{center}
  \includegraphics{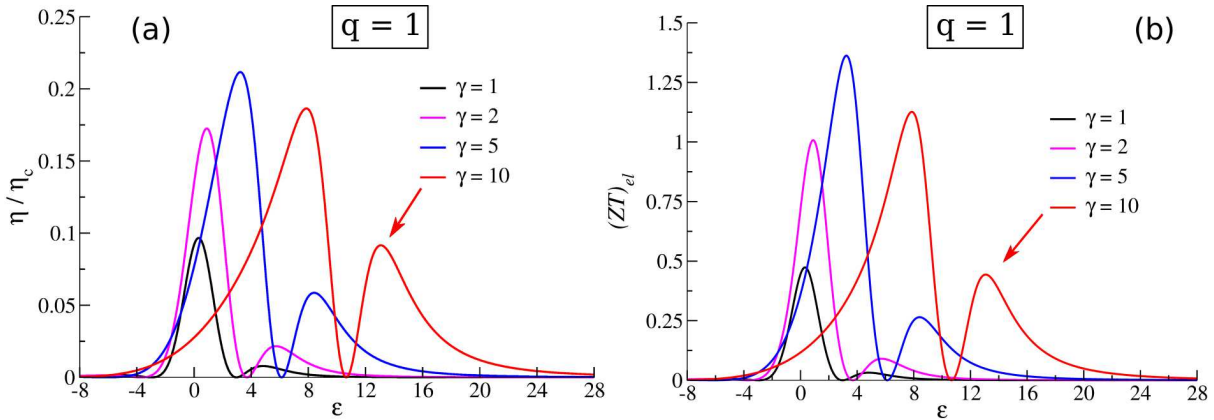}
	\end{center}
		\vspace{-0.8cm}
	\caption{Universal  curves for: (a) the efficiency $\eta/\eta_c$  and (b) the figure of merit  $(ZT)_{el}$, of the  thermal machine with Fano lineshape 
   as a function of the dimensionless energy parameter $\varepsilon= (E_d-\mu)/k_BT$,   for fixed values  
   of the broadening  parameter $\gamma= \Gamma/k_BT$. The asymmetry parameter  $q$ has 
   been set equal to 1.}
\end{figure}    

    We begin with the discussion of the behavior of the   
    efficiency $\eta/\eta_c$, and we report in Fig.8a the family of universal 
   curves   for the efficiency of the  thermal machine with Fano lineshape 
   as a function of the dimensionless parameter $\varepsilon$,
   for fixed values  of the broadening  parameter $\gamma$;  in these 
   simulations we set the value  of the $q$ parameter equal to unity. 
   We notice that  the efficiency   is not symmetric   with respect 
    to  $\varepsilon$, and goes to zero for large $|\varepsilon|$.
     A comparison of Fig.8a (corresponding to $q=1$) 
    and  Fig.5a (corresponding to $q=0$) shows that the efficiencies 
    for $\varepsilon >0$  for Fano lineshapes are  enhanced with respect to the 
    efficiencies for $\varepsilon >0$ of the antiresonance.  In the case  of Fano lineshapes 
    the asymmetry parameter  adds further flexibility   to the engineering of molecular thermal machines.

   We pass now  to the Seebeck  coefficient. In Fig.9 we report the thermopower for the Fano transmission lineshape for fixed value $\gamma=1$ and $q=0, \,\pm 1, \, \pm 2$. As expected  the curves with parameters $\pm q$ exhibit inversion symmetry  with respect to the origin  of the $\varepsilon$ variable. It is important  to notice that the curves with $q=0, \,\pm 1, \, \pm 2$ (as well other values not reported) show a substantial increase of  the absolute value of the Seebeck thermopower, compared with the curve $q=0$ of the antiresonant structure and already discussed in Fig.6.
   
      \begin{figure}[h]
	\begin{center}
  \includegraphics{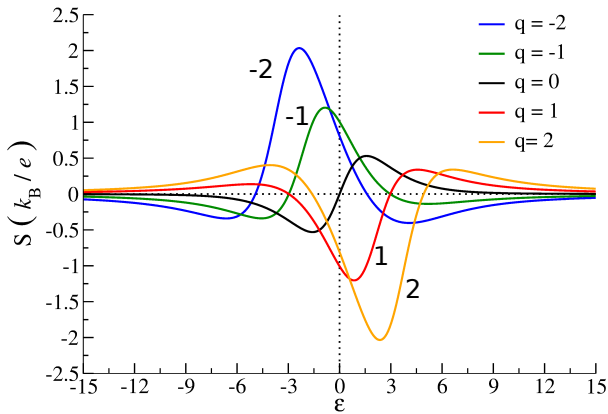}
	\end{center}
		\vspace{-0.8cm}
	\caption{Thermoelectric 
   power (in units $k_B/e$) for Fano transmission functions, versus  
   the $\varepsilon= (E_d-\mu)/k_BT$ energy parameter, for different values of the asymmetry parameter $q$.
   The broadening  parameter $\gamma= \Gamma/k_BT$ has been chosen equal to unity.
   }
\end{figure}    

InFig.10, i we report the thermoelectric power of  Fano transmission functions for fixed values of $\gamma$, and values $q=1$ and $q=5$ of the asymmetry parameter. From Fig.10 it is seen  that the Seebeck coefficient assumes
   absolute values  of the order of $k_B/e$  or more
   as $\gamma$ increases; the same occurs  for increasing values of $q$,  
  as it can also be 
   confirmed  by appropriate manipulations  of the trigamma function.  Thus 
   from a qualitative point of view,  the analysis  of  the Fano structure
   hints  the possibility of good thermoelectric devices with  high 
   efficiency and Seebeck thermopower, and relatively large broadening. 
   In summary,  the  link between good efficiency and  small  broadening 
   of Lorentzian lineshapes  is broken to some extent in antiresonances, 
   and further relaxed  for  Fano-like transmission lineshapes.
   
      \begin{figure}[h]
	\begin{center}
  \includegraphics{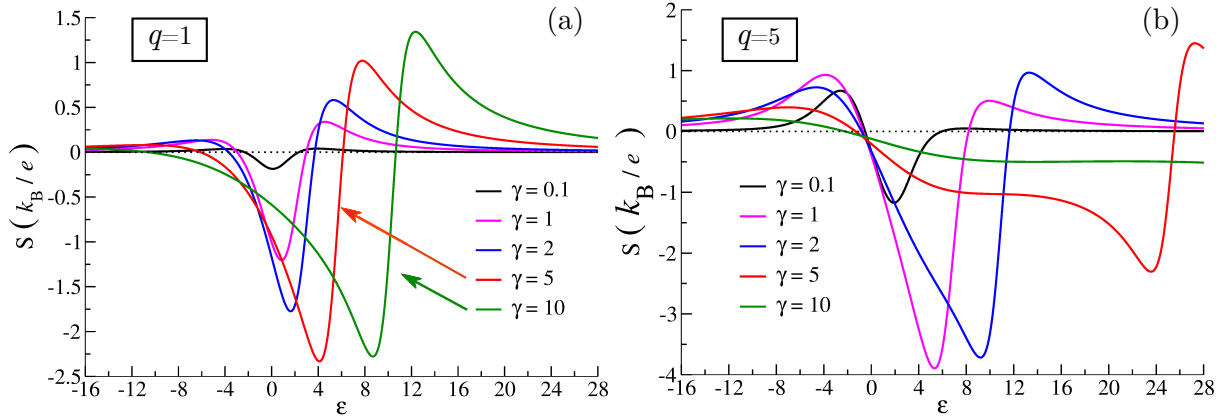}
	\end{center}
		\vspace{-0.8cm}
	\caption{(a) Thermoelectric 
   power (in units $k_B/e$) for Fano transmission functions, versus  
   the energy parameter $\varepsilon= (E_d-\mu)/k_BT$, for  fixed values  of the broadening  
   parameter $\gamma= \Gamma/k_BT$. The asymmetry parameter  $q$ has 
   been set equal to 1  in panel (a)  and equal to 5 in panel (b).}
   \end{figure}    


      \section{Conclusions}

     This article addresses quantum transport 
    through nanoscale  thermoelectric
    devices, and   discuss  the general 
    equations controlling  the electric charge current,  heat currents, and 
    efficiency of energy transmutation in steady conditions  in the linear regime. 
    With focus in the parameter domain where the electron system  acts  
    as a  molecular power-generator,  we  provide  the expressions 
    of optimal efficiency,  electric and thermal conductance, Lorenz number and 
    power-output of the device. The treatment is fully analytic  and presented in terms of   
     trigamma functions and Bernoulli numbers.
    
    The general concepts are put at work in  paradigmatic devices  
    with Lorentzian  resonances and anti-resonances transmission 
   functions. A most important feature of this investigation is the emergency 
   of the  complementary roles of peaked and valleyed structures: 
    in the former the most interesting region of application involves 
    long-lived electron states $(\gamma \ll 1)$, while in the latter it involves 
     $\gamma \gg 1$ structures. 
    The simulations are then extended  to the paradigmatic case  of   Fano  
    transmission functions, that encompass peaked and valleyed regions.
    In the case  of Fano lineshapes,  the role of the asymmetry  parameter can 
    be exploited to widen the region of good performance of the devices, 
    and to  add further flexibility 
    to the engineering of molecular thermal machines. 
           
   The procedures elaborated in this article can be extended  
   to non-linear  situations, as well as to  systems  with broken time-reversal 
   symmetry.  Within the framework of the non-equilibrium Keldysh formalism, 
   the approach can  be  generalized to handle interacting  quantum systems 
  and in particular  electron-phonon interactions, which have been 
  so fruitfully explored in the lowest order approximation. In all these 
  variegated subjects, the approaches elaborated in  this work  can  be of 
  help for developing protocols and in depth   understanding of the non-equilibrium processes accompanying  
  charge and heat currents, and efficiency of energy 
   transmutation in nanoscale devices.

 \vspace{0.5cm}
 
\noindent  {\bf Appendix A. Some integrals  for the analytic 
               treatment of thermoelectricity with Fano lineshape transmission} 
      \setcounter{equation} {0}
       \numberwithin{equation}{section}
       \renewcommand\theequation{A\arabic{equation}}
  
  \vspace{0.5cm}
  
\noindent   In the treatment of thermoelectric effects in nanoscale systems 
   with Fano or  Fano-like lineshapes in the   
  linear regime, we have to consider  integrals of the type
          \begin{equation}
                    I_n(w) =  i   \int_{-\infty}^{+\infty}  dz \,  
                \frac{z^n}{ z- w } \, (- \frac{\partial f}{\partial z}) \ \  
	           (n=0,1,2, \ldots) \quad   f(z) = \frac{1}{e^{z} +1} \ \ , \ \ 
	                      {\rm Im} \, w <0 \ ,  
            \end{equation} 
    where $f(z)$ is the Fermi function (with unitary thermal energy and 
    zero chemical potential), and $w$ is a complex variable located 
    in the lower half part of the complex plane. The purpose of this Appendix 
    is to provide an analytic expression of the $I_n(w)$-integrals, which 
    are just  the key ingredient for the calculation of the   thermoelelctric 
    parameters. First, we show that $I_0(w)$ can be calculated 
    analytically with the trigamma function.  Next, by virtue of  
    recurrence relations, we express any $I_n(w)$ with  $n=1,2,3, \ldots$
    in terms of $I_0(w)$. 
    
    \vspace{0.5cm}     
   \noindent   {\bf Analytic evaluation of $I_0$}  
      \vspace{0.5cm}
       
   \noindent  The auxiliary  integral $I_0(w)$,  according to   Eq.(A1),  reads
        \begin{equation}
                    I_0(w) =  i   \int_{-\infty}^{+\infty}  dz \,    \frac{1}{ z- w } 
                    \, (- \frac{\partial f}{\partial z})   \quad , \quad {\rm Im}\, w <0 
                    \quad, \quad   f(z) = \frac{1}{e^{z} +1}  \ .  
            \end{equation}  
      For the analytic evaluation of  $I_0$   we exploit the multipole  
      series expansion of the derivative  of the Fermi 
      distribution function.

    The Fermi-Dirac distribution function can be expanded in the series      
        \begin{equation*}
            f(z) = \frac{1}{e^{z} +1} \equiv \frac{1}{2} 
                 - \sum_{n=-\infty}^{+\infty} \, \frac{1}{\, z -  (n+1/2)2 \pi i \, } \ ;
         \end{equation*} 
  differentiation of  both members of the above equation gives:
       \begin{equation*}
         \frac{\partial f}{\partial z} = 
              \sum_{n=-\infty}^{+\infty} \,  \frac{1}{ \, [ z - (n+1/2) 2\pi i ]^{2} \, } \ .
      \end{equation*} 
   The Fermi function is represented by a ladder of poles of the first
   order along the imaginary axis with steps of $2\pi i$; the derivative of 
  the Fermi function  is represented  by a ladder of second 
  order poles along the imaginary axis with steps of $2\pi i$.

  With the multipole expansion of the derivative of the Fermi 
  function,  the integral $I_0(w)$ defined by Eq.(A2) becomes
          \begin{equation}
          I_0(w) =  i \sum_{n=-\infty}^{+\infty} \,  
                   \int_{-\infty}^{+\infty}  dz \,   
                          \frac{1}{z - w} \cdot \frac{-1}{ [z - (n+1/2)2 \pi i  ]^{2} }  
                          \qquad  ( {\rm Im} \, w <0 )\ .
         \end{equation} 
   The pole of the first function in the  integrand occurs 
   at $z= w$, which is in the lower part  of the 
   complex plane; thus we close the integration path on the upper part 
   of the complex plane. The singularities of the integrand  in the upper
   part of the complex plane are represented by poles of the second order,
   placed at the points of the imaginary axis
         \begin{equation*}  
                z = z_{n} \equiv   (n+\frac{1}{2}) 2 \pi i     \quad   (n=0,1,2,...) \ ;
       \end{equation*}
   for the residues, we need the derivative
          \begin{equation*}
               \frac{d \ }{dz} \, \left[ \frac{1}{z - w} \right] = \frac{-1} { (z -w)^{2}} \ .
         \end{equation*}    
   Due to the presence of the above poles of second order,  the integral
   in Eq.(A3)  becomes       
        \begin{eqnarray*}
           I_{0}(w)  &=& i  \, \sum_{n=0}^{+\infty} \, 2\pi i \, 
                          \frac{-1}{( z_{n} - w)^{2}} (-1)
                 = i \,  \sum_{n=0}^{+\infty}  \ 
                   \frac{2\pi i}{ \left[ ( n + 1/2)2\pi i  - w \right]^2 }  
                                         \\[2mm]
                   &=& \frac{i}{2\pi i} \sum_{n=0}^{+\infty}  \ 
                   \frac{1}{  (n + 1/2  +i w/ 2\pi  )^2 }  \ .
         \end{eqnarray*}
     It follows
          \begin{equation} 
                 I_{0}(w) = \frac{1}{2\pi} \Psi_t \, (\frac{1}{2} + \frac{i w}{2\pi})  \ ,
          \end{equation} 
     where
          \begin{equation} 
                    \Psi_{t}(z) = \sum_{n=0}^{\infty} 
                                  \, \frac{1}{(z+ n)^{2}}  
           \end{equation} 
       is the trigamma function. For details on the digamma, trigamma and 
       poligamma functions see for instance Ref.\onlinecite{ABRA}

       \vspace{0.5cm}    
   \noindent   {\bf Analytic expression of $I_n(w)$  
                                               with recursion relations}
      \vspace{0.5cm} 
                                                   
  \noindent   Having established the analytic expression of the $I_0(w)$ function,
     we pass now to the  analytic expressions of $I_n(w) \ (n\ge1)$  
     exploiting appropriate recursion relations.
     We start from the identity
         \begin{equation*}
           \frac{z^n}{z-w} \equiv z^{n-1} + w\, \frac{x^{n-1}}{z-w} 
               \qquad n\ge 1 \ .  
          \end{equation*}
   Multiplying all members of the above identity by $i(-\partial f)/(\partial z)$, 
   and integrating over $z$ on the real line, we obtain 
          \begin{equation}
         i \int_{-\infty}^{+\infty}  dz \, \frac{z^n}{z-w} \, (- \frac{\partial f}{\partial z})
                =  i   \int_{-\infty}^{+\infty}  dz \, z^{n-1} \, (- \frac{\partial f}{\partial z})
                + i w   \int_{-\infty}^{+\infty}  dz \, \frac{z^{n-1}}{z-w} 
                        \, (- \frac{\partial f}{\partial z}) \ . 
            \end{equation} 
    The  integrals  appearing at the beginning of  the right hand  side 
    of Eq.(A7) are closely related to the well known Bernoulli numbers, 
    frequently  encountered in several fields of condensed matter physics. 
    It holds
                 \begin{equation}
           \int_{-\infty}^{+\infty}  dz \, z^m \, (- \frac{\partial f}{\partial z})
                =  b_m \qquad m=0,1,2, \ldots  
            \end{equation} 
     where the first few Bernoulli-like numbers $b_n$ are  
            \begin{equation}
                 b_0 = 1 \ , \ b_1 = 0 \ , \ b_2 = \frac{\pi^2}{3} \ , \ b_3 = 0 \ ,
                          \   b_4 = \frac{7 \pi^4}{15} \ , \  b_5 = 0 , \, \ b_6 = \frac{31\pi^6}{21} \ ,\ \ldots
             \end{equation}  
    The Bernoulli-like number of odd  order are all zero for 
    symmetry reasons.

   The structure of Eq.(A6) defines the recursion relation
            \begin{equation} \boxed{ \
                  I_n(w) = i  b_{n-1} + w \, I_{n-1}(w)  \ }   \qquad n\ge1   \ .
            \end{equation} 
    Thus the knowledge of $I_0(w)$ entails 
    the knowledge of all the  auxiliary integrals $I_n(w)$. 
   The first  few $I_n(w)$   for $n=0,1,2,3,4$ in terms of $I_0(w)$ read  
        \begin{eqnarray}
                 I_{0}(w) &=&   \frac{1}{2\pi} 
                            \,  \Psi_t(\dfrac{1}{2} + \dfrac{i  w}{2\pi})
                                     \qquad  ( {\rm Im} \, w <0 )
                                  \nonumber   \\[2mm]
                I_{1}(w) &=& i + w I_{0}(w)
                                   \nonumber\\[2mm] 
                  I_{2}(w) &=& iw + w^2 I_{0}(w)
                                      \nonumber\\[2mm]
                    I_{3}(w) &=&ib_2 + iw^2 + w^3 I_{0}(w)  
                                      \nonumber   \\[2mm]                                                    
                   I_{4}(w) &=&iwb_2 + iw^3 + w^4 I_{0}(w) \ .
          \end{eqnarray}  
    By virtue of the analytic  expressions  summarized in Eqs.(A10), the
    thermoelectric parameters and transport    of nanoscale devices 
    with Fano lineshapes, Lorentzian lineshapes and antiresonance 
    lineshapes can be elaborated in analytic forms, particularly suitable for 
    a deeper description and  investigation of the variety of quantum 
    physical effects  emerging  in  nanostructures.

    \vspace{0.5cm} 
\noindent {\bf Appendix B. Optimal efficiency of nanoscale devices} 
      \setcounter{equation} {0}
       \numberwithin{equation}{section}
       \renewcommand\theequation{B\arabic{equation}}
  
      \vspace{0.5cm} 
 \noindent   In this Appendix we present a simple and self-contained elaboration  
    of the optimal efficiency expression for nanoscale devices. This is useful not only for 
    a deeper investigation  of the transport properties in nanostructures,
    but also because most theoretical treatments are   spread, 
    not to say entangled,  in a variety of articles and other sources.

     We start from the expression  of the efficiency parameter  
    given by Eq.(12d) of the main text; namely:
            \begin{equation}
             \eta =  \frac{ \  - K_0 \, (\Delta V)^2  
             \ + \ \dfrac{k_BT}{e}  K_1 \, \Delta V \, \dfrac{\Delta T}{ T} }
                         {  - \dfrac{k_BT}{e}  K_1\,  \Delta V 
                 + \dfrac{ k_B^2 T^2}{ e^2} K_2 \, \dfrac{\Delta T }{ T} }  \ .
           \end{equation} 
     We focus on the parameter domain of Eq.(18a), where the power 
    output  is positive, and  write
           \begin{equation}
                \Delta V  =  x \,  \frac{K_1}{K_0} \, \frac{k_BT}{e} 
                   \, \frac{\Delta T}{T} 
                      \qquad with \qquad  0 < x < 1 \ ,
            \end{equation} 
    where $x$ is a dimensionless parameter confined in the interval $[0,1]$.
    Replacement of Eq.(B2) into Eq.(B1) gives 
             \begin{eqnarray}
                 \eta  &=&  \frac{  -  K_0 \,  x^2  \, 
                                  \dfrac{ K_1^2}{K_0^2} 
                           \left(\dfrac{\Delta T}{T}\right)^2
                                   +  K_1\, x \,  \dfrac{ K_1}{K_0} 
                           \left(\dfrac{\Delta T}{T}\right)^2 }
                         {  -  K_1\, x \,   \dfrac{ K_1}{K_0} 
                           \dfrac{\Delta T}{T} +  K_2  \,  \dfrac{\Delta T}{T}    } 
                                            \nonumber  \\[2mm]
                  & = & \frac{  -  x^2  \,  \dfrac{ K_1^2}{K_0} 
                            \dfrac{\Delta T}{T}
                                   +  x \,  \dfrac{ K_1^2}{K_0} 
                               \dfrac{\Delta T}{T}   }
                         {  -  x \,  \dfrac{ K_1^2}{K_0} 
                            +  K_2 \dfrac{ K_0}{K_1^2}  \, 
                           \dfrac{ K_1^2}{K_0}    } 
                              \quad [set \ p = \frac{K_1^2}{K_0K_2}  \quad with 
                              \quad  0 \le p \le 1 ] \, . \quad
           \end{eqnarray} 
     The efficiency of a device, characterized by a specific 
     parameter $p\, (<1)$ becomes
             \begin{equation}
                 \frac{\eta}{\eta_c} =  \frac{x - x^2}{1/p -x}
                      = p \, \frac{x - x^2}{1 -  p x}      \equiv f(x; p)  
                                      \qquad with         \qquad   0 < x < 1 \ ,
             \end{equation}  
      where the function $ f(x;p)$ provides the  efficiency of 
      the device of parameter $p$, compared with the efficiency of the Carnot 
    machine operating between the same temperatures of the two reservoirs.

    The maximum value  of the efficiency function occurs for
            \begin{equation*}
                  \frac{d f(x;p)}{dx} = p \, \frac{(1 - 2x) (1 - px) + p( x -x^2)}{(1-px)^2}
                         =p \, \frac{ px^2 - 2x + 1}{(1-px)^2} = 0 \ .
             \end{equation*}  
     The solution in the interval $[0,1]$ of interest is 
             \begin{equation*}
                    x_0 = \frac{1}{p} \left[ 1 - \sqrt{1 - p} \, \right ]    \ .
              \end{equation*}
      The optimized  value of the efficiency function becomes
            \begin{eqnarray*}
               f(x_0;p) &=& p \, x_0 \,  \frac{1 - x_0}{1 -p \,x_0}
                   = (1 - \sqrt{ 1 - p}) \cdot  \left[ 1 - \frac{1}{p}(1- \sqrt{ 1 - p}) \, \right]
                   \cdot \frac{1} { \sqrt{ 1 - p}}   
                                                 \\[2mm]
                  &=& (1 - \sqrt{ 1 - p}) \cdot \frac{ 
                                    p-1 +  \sqrt{ 1 - p}}{p} \cdot \frac{1} { \sqrt{ 1 - p}}  
                            = \frac{1}{p} \left( 1 - \sqrt{ 1 - p} \right)^2\ .
            \end{eqnarray*}   
        In summary:
             \begin{equation} 
                  \frac{\eta}{\eta_c} = \frac{1}{p}  \left( 1 - \sqrt{ 1 - p} \right)^2    \ .
             \end{equation} 
     It is almost superfluous to verify that the optimal efficiency  of the 
     device  is smaller than   the Carnot cycle efficiency. This becomes 
     even more apparent  using in Eq.(B5) the identity
                  \begin{equation*}
                 p \equiv \left[ 1- \sqrt{ 1 - p} \right] \left[ 1+ \sqrt{ 1 - p} \right] \ .
                 \end{equation*}
      We obtain the self-explaining relation
                    \begin{equation}
                         \frac{\eta}{\eta_c} =  \frac{  1 - \sqrt{ 1 - p}  }
                                        {  1 + \sqrt{ 1 - p}  }  \ ,       
                   \end{equation}   
      that makes even more evident the physical meaning  of the 
      performance parameter   defined in this article. It is apparent that  the 
      efficiency function takes its maximum  value $1$ 
      for $p=1$, and decreases monotonically to zero for decreasing
      values of $p$ performance parameter. In the literature alternative 
      more or less popular   performance parameters,  or figure-of-merit 
      parameters, are  used to characterize thermoelectric devices.
       In this article we stick to our elaboration because of its 
      simplicity and  fully self-contained derivation. 
      
     Good thermoelectric devices should have the 
     performance $p$-parameter as near as possible  to unity, preferably
    in the  range $p \in [0.8 -   1]$ or so, corresponding to efficiency  
    from $25 \%$  to $100 \%$, relative to the Carnot cycle. This range of 
    values is   argued to be competitive with conventional gas-liquid 
    compressor-expansion motors.

      Similar considerations can be worked out  to establish the parameter 
    region where  the molecular device  acts as a refrigerator, with heat 
    current flowing from the cold reservoir to the hot one with 
    absorption of  external work.

         \
         
         {\bf Acknowledgments}
The authors acknowledge the "IT center" of the University of Pisa for the computational support. We acknowledge also the allocation of computer resources from CINECA, ISCRA C Projects HP10C6H6O1, HP10CAI9PV.



\end{document}